\documentclass[10pt]{article}

\usepackage[pdftex]{graphicx}
\graphicspath{{./pdf/}}

\usepackage{amsmath} \allowdisplaybreaks
\usepackage{amssymb}
\usepackage{moreverb}
\usepackage{metalogo}

\usepackage{csagh}

\usepackage{float}

\usepackage{lineno}
\nolinenumbers

\begin{document}

Computer Science • 20(4) 2019 \quad \quad \quad \quad \quad \quad https://doi.org/10.7494/csci.2019.20.4.3376

\begin{opening}

\title{Track finding with Deep Neural Networks}

\author[Institute of Nuclear Physics PAN, Krakow, Poland,
         \URL{http://www.ifj.edu.pl},
         e-mail: \URL{marcin.kucharczyk@ifj.edu.pl}]
       {Marcin Kucharczyk}
       
\author[Institute of Nuclear Physics PAN, Krakow, Poland,
\URL{http://www.ifj.edu.pl},
e-mail: \URL{marcin.wolter@ifj.edu.pl}]
{Marcin Wolter}

\date{Krakow, 20.03.2019}

\begin{abstract}

High-energy physics experiments require fast and efficient methods for reconstructing the tracks of charged particles. The commonly used algorithms are sequential, and the required CPU power increases rapidly with the number of tracks.
Neural networks can speed up the process due to their capability of modeling complex non-linear data dependencies and finding all tracks in parallel.

In this paper, we describe the application of a deep neural network for reconstructing straight tracks in a toy two-dimensional model.  
It is planned to apply this method to the experimental data obtained by the MUonE experiment at CERN.

\end{abstract}


\end{opening}

\section{Introduction}

Particle detectors used in high-energy physics (HEP) require the efficient and fast reconstruction of charged particle tracks. Effective tracking algorithms have been used for years in HEP experiments (for example, the Kalman Filter~\cite{fruhwirth1987application}), but they are sequential - the required CPU power increases rapidly with track density. Machine learning algorithms, especially Deep Neural Networks (DNN), can speed up the reconstruction process thanks to their capability to model complex non-linear data dependencies and parallelize easily.
Besides the reduction of CPU consumption, it is important how the efficiency and precision of DNN reconstruction compares with standard track reconstruction methods in the use of a neural network 

Our goal is to check the potential and eventually apply the track-finding techniques based on machine learning. They are planned to be applied to the experimental testbeam data taken in 2018 by the MUonE experiment~\cite{MUonE}, which will be operating at the SPS accelerator at CERN. The experiment is supposed to measure a hadronic contribution to the anomalous muon magnetic moment in order to increase the sensitivity to the potential new physics phenomena. This might cause a discrepancy with respect to the standard model predictions, which would be a clear sign of new physics. For the MUonE experiment, the tracking is essential; deep neural networks may provide fast and efficient pattern recognition, which is the most crucial step in the track-reconstruction procedure.

\section{Track reconstruction}
\label{reco}

Our aim is to test whether we are able to reconstruct a particle track on a simple toy example without using sequential tracking algorithms while reconstructing the track parameters all at once using the deep neural network. 
The previously trained neural network is fast and does not need much CPU power; however, the training of a DNN requires quite a bit of CPU power. For training bigger networks, graphics processor units (GPUs) are needed. Also, the amount of available memory might be a problem.  
To ensure fast training on limited resources, a two-dimensional 28$\times$28 pixel toy model with straight particle tracks has been developed.
A random noise and pixel inefficiency has been added to make the model more realistic. 

The input to our neural network is a two-dimensional pixel map representing the hits in the detector; no data preprocessing was done.

\subsection{Reconstruction of single track}

\begin{figure}[!ht]
	\centering
	\includegraphics[scale=.45]{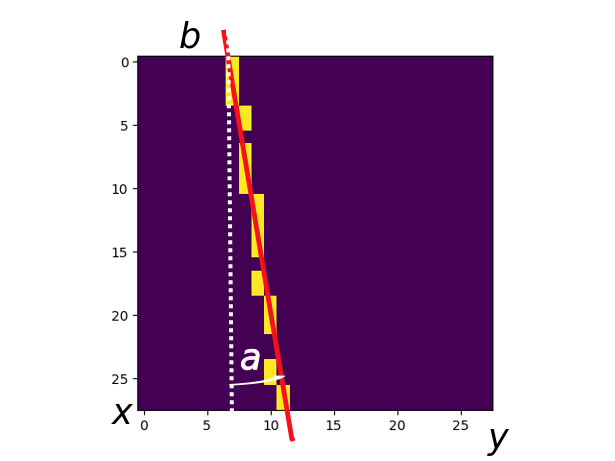}	
	\caption{Single track event. Highlighted pixels are actual hits: red line represents true track; \emph{a} represents track slope; and \emph{b} represents intercept.  
	}
	\label{fig_slope}
\end{figure}

A straight track in two-dimensional space can be described by the equation representing a straight line (i.e., $y=ax+b$, with slope $a$ and intercept $b$ [see Figure~\ref{fig_slope}]), so the neural network should return these two parameters. The parameters of the generated tracks were restricted in such a way that the track always crosses the upper and lower edges of the detector. For the described analysis, the training and test samples of 20,000 events were used for both single and multiple track events.

The neural network is trained to perform the regression using the entire 28$\times$28-pixel matrix as an input and returning the track parameters. Since this task is similar to the image processing, we opted to use the convolutional neural network (CNN)~\cite{lecun2015deep}, which proved to be very powerful in the image-recognition tasks; this is routinely used in many applications.

Traditional multilayer perceptron (MLP) models have been also used for image recognition. However, due to the full connectivity between nodes, they suffer from the curse of dimensionality and do not scale well with higher-resolution images. Also, the feature extraction might be different in different regions of the image.

\subsubsection*{Convolutional layer}

In the convolutional layer, the layer's parameters consist of a set of filters that have a relatively small receptive field. The filters are trained to activate when some specific type of feature is detected; this allows one single filter to find given features in the entire input image.
A stack of convolutional layers works as an auto-encoder, extracting more complex features in each layer.

%

\subsubsection*{Pooling layer}

A pooling layer reduces the spatial size of the representation, as the exact location of a feature is less important than its approximate position relative to other features. 
The use of pooling layers helps to control overfitting and reduce the amount of computation time. 


\subsubsection*{Dropout layer}

In the dropout approach at each training stage, a set of nodes randomly selected with a given probability are "dropped out" from the network. A reduced network is left and is trained during this stage. After this training stage, the removed nodes are returned to the net with their original weights. 
This approach introduces some random noise to the network and decreases overtraining.

%
%

\subsubsection*{Network architecture}

\begin{figure}[!ht]
	\centering
	\includegraphics[width=\textwidth]{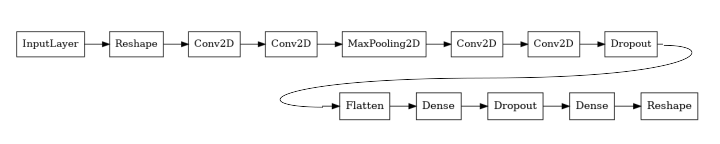}	
	\caption{Convolutional deep neural network used to reconstruct track parameters for single track events.}
	\label{fig:CNN}
\end{figure}

The CNN network is implemented using the KERAS~\cite{chollet2015keras} library with tensorflow~\cite{abadi2016tensorflow} as a backend.  
The network used to reconstruct single tracks consists of two convolutional layers (\emph{Conv2D}) with 8 convolution windows with a size of 3$\times$3 followed by the \emph{MaxPooling} layer with a pooling size of 2$\times$2. This is again followed by another two convolutional layers, each having 32 windows of a size of 3$\times$3. The single \emph{Dropout} layer with a 0.25 dropout fraction suppresses the overtraining, and the final regression is performed in the \emph{Dense} layer with 400 nodes and the rectified linear unit (ReLU) activation function.
This dense layer is followed by another \emph{Dropout} layer with a 0.5 dropout fraction and the final \emph{Dense} output layer with two nodes, which returns the reconstructed track parameters. This layer has a hyperbolic tangent activation function.
The neural network (presented in Figure~\ref{fig:CNN}) has more than 800,000 trainable parameters in total.

\begin{figure}[!ht]
	\centering
	\includegraphics[width=0.6\textwidth]{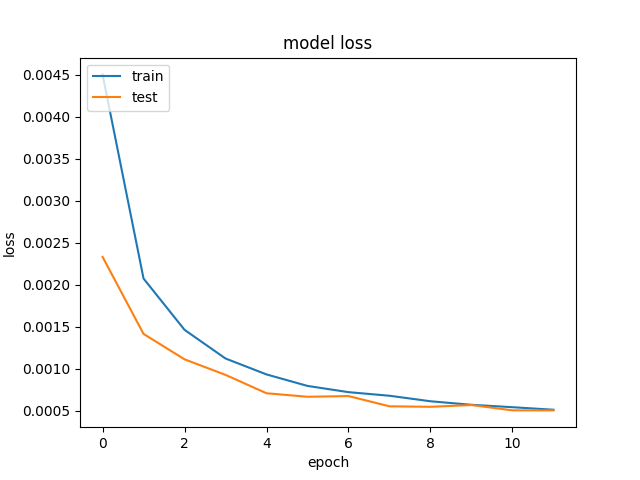}	
	\caption{Evolution of loss function during training for single track events.}
	\label{fig:CNN_loss}
\end{figure}

The network is trained using the Adam optimizer~\cite{kingma2014adam}, and the mean squared error is used as a loss function. The training is performed for 12 epochs (with a batch size of 128), which is sufficient to reach a plateau of the loss function. The evolution of the loss function during training is shown in Figure~\ref{fig:CNN_loss}.

The model with the convolutional neural network and long/short-term memory (LSTM) layer is based on the research presented by the HEP.TrkX project~\cite{heptrkx} in Refs.~\cite{farrell2017hep, farrell2018novel}. 

\subsection{Reconstruction of multiple tracks}

A network described above is capable of reconstructing events with a single track. For events with multiple tracks, an additional layer is needed; this transforms the output of the neural network into a sequence of track parameters. The number of reconstructed tracks might vary from event to event.

\subsubsection{Long/Short-Term Memory layer} 

The multiple track finding problem might be treated in a similar fashion to the image-captioning, where the descriptions of the tracks (i.e., track parameters) are analogous to the text captions assigned to the various patterns seen in the image \cite{vinyals2015show}.
For this purpose, the long/short-term memory (LSTM) \cite{hochreiter1997long} layer is used. 
LSTM is frequently used to process time signals; i.e., a flow of sequential inputs. In the case of track reconstruction, the CNN network sequentially reconstructs consecutive tracks, which are the sequential input for the LSTM layer.

\subsubsection*{Network architecture}

\begin{figure}[!ht]
	\centering
	\includegraphics[width=\textwidth]{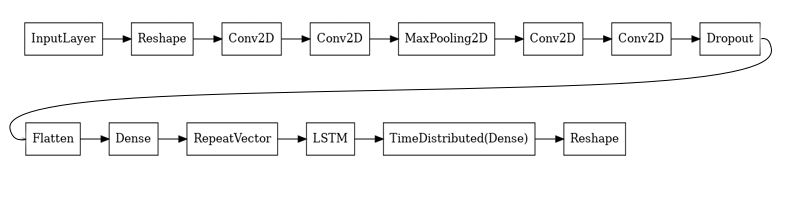}	
	\caption{Convolutional deep neural network with convolutional layers followed by dense and LSTM layers. Network is trained to reconstruct track parameters for multiple track events.
	}
	\label{fig_model}
\end{figure}

While reconstructing multiple track events, the convolutional neural network (which is identical to the one used for single track events) extracts all relevant features from the input image. The \emph{LSTM} layer stacked on top of the convolutional and dense layers emits the track parameters for each track in a sequence, thus finding the parameters of consecutive tracks. The diagram of a complete network is shown in Figure~\ref{fig_model}.

The first part of the network (i.e., the convolutional layers and the first dense layer) remain unchanged. The \emph{LSTM} layer with 400 units, the hyperbolic tangent activation function, and the sigmoid recurrent activation function is used. This is followed by a \emph{Dense} with two units and the linear activation function. It is wrapped in the \emph{TimeDistributed} wrapper, which applies this dense layer to each temporal slice of an input.
	
The test of multiple track reconstruction is performed for events with exactly three simulated tracks. Therefore, the last layer of the network is a \emph{Reshape} layer, which forms the output into a 3$\times$2 array containing the parameters of the three tracks. During the training process, the network output is compared to the true parameters of the three simulated tracks. 

\begin{figure}[!ht]
	\centering
	\includegraphics[width=0.6\textwidth]{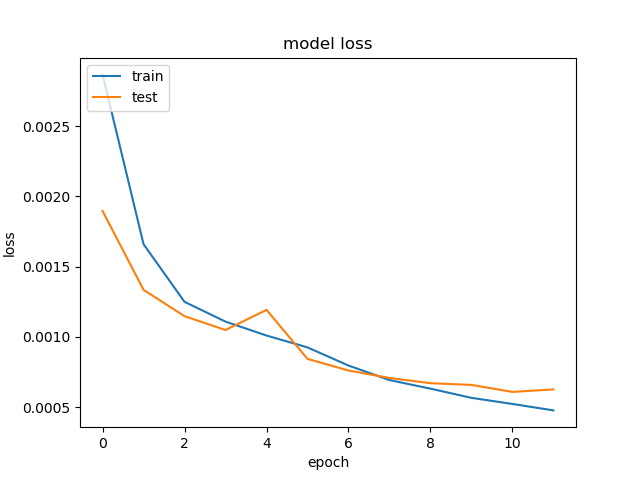}	
	\caption{Evolution of loss function during training for multiple track events.}
	\label{fig:LSTM_loss}
\end{figure}

The neural network described above has more than two~million trainable parameters.
The network is trained in the same manner as with the single tracks. The evolution of the loss function during training is shown in Figure~\ref{fig:LSTM_loss}. The training should be stopped at this point to avoid overtraining.

\subsubsection*{Track fitting}  

After performing the neural network reconstruction, the hits close to the track returned by the network are used to perform the straight line fit. In order to include only relevant hits to the fit, a cut-off of five pixels around the DNN track (Euclidian distance) is chosen. Such a cut-off ensures a reasonable balance between including many noisy hits and not losing too many hits originating from a real track, since the reconstruction performed by a neural network is never perfect. The accuracy of a hit is limited to the pixel size, so all of the hits are given the same error equal to one for the fitting.

All of these hits are used to fit a track using the robust fit~\cite{motulsky2006detecting} implemented in the ROOT \cite{Brun:1997pa} framework. The robust fit allows us to drop the outliers and redo the fit with fewer hits. The robust fit is performed with a limit of at least 70\% of the original input hits to be used in the final fit.
The fit (performed after the neural network pattern recognition) is implemented to improve the resolution of the track reconstruction done by a neural network. The fitting algorithm in the ROOT package is highly optimized, so the robust fit uses only about 5\%~CPU more than the standard fitting procedure does.

\section{Results}

The track reconstruction was first performed for events with a single track. In the next step, the analysis was repeated for events with three tracks. 

\subsection{Single-track events}

In the first step of our analyses, we tested the performance of the convolutional neural network without an LSTM layer on single-track events. The pixel efficiency was set to 70\%, and the noise varied within a range of 0 to 30\%. 
The noise level is defined as the probability of a single pixel not belonging to the track to generate a false signal. 
It should be emphasized that 30\% noise is quite extreme and unrealistic for real detectors. It is only included here to test the robustness of the pattern recognition based on neural networks.

\begin{figure}[!ht]
	\centering
	\includegraphics[clip,trim={80px 0 100px 0},width=0.40\textwidth]{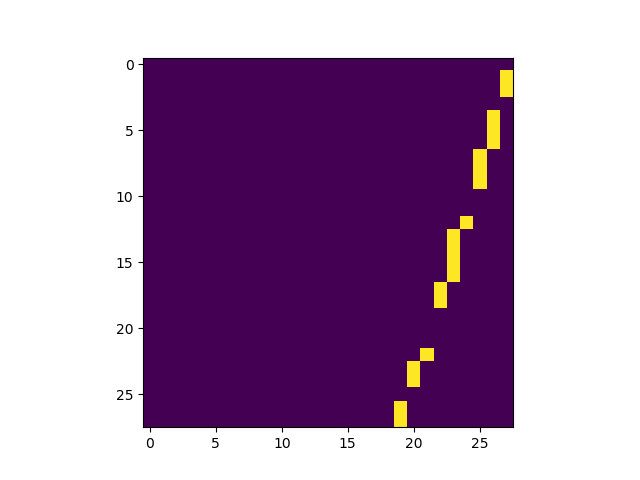}	
	\includegraphics[clip,trim={80px 0 100px 0},width=0.40\textwidth]{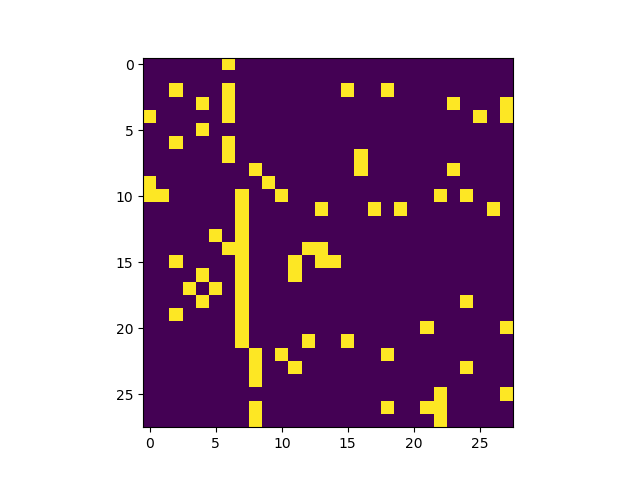}	
	\includegraphics[clip,trim={80px 0 100px 0},width=0.40\textwidth]{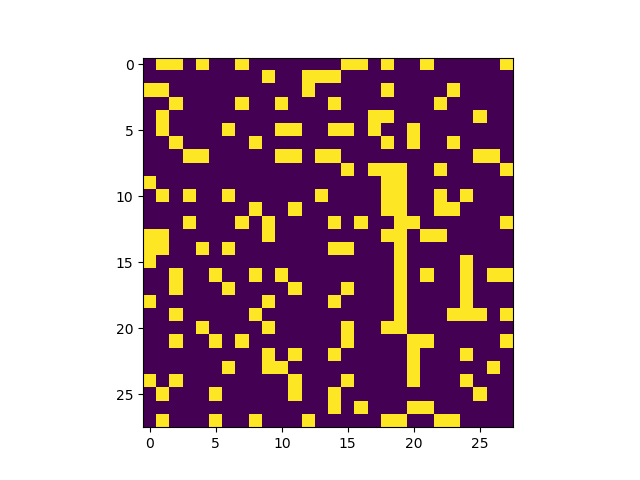}	
	\includegraphics[clip,trim={80px 0 100px 0},width=0.40\textwidth]{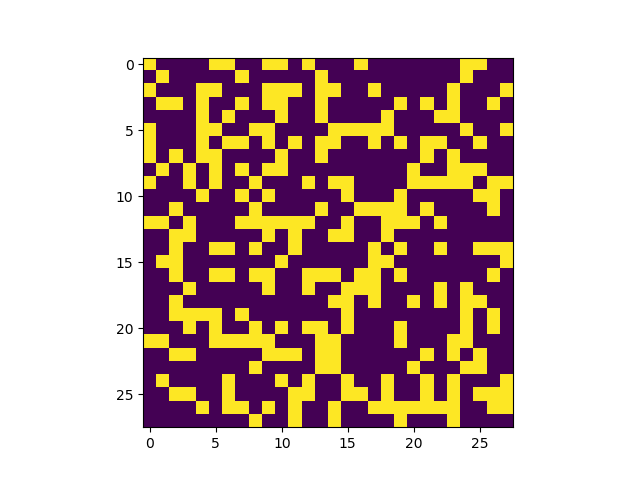}	
	\caption{Single-track events with pixel efficiency of 70\% and noise levels of 0 (upper left), 10\% (upper right), 20\% (bottom left), and 30\% (bottom right). Units are pixel numbers. }
	\label{fig_single_event}
\end{figure}

The hit positions are not smeared, but the hit accuracy is limited to the size of a pixel. 
The example events with various noise levels are shown in Figure~\ref{fig_single_event}.

\begin{figure}[!htbp]
	\centering
	\includegraphics[width=1.0\textwidth]{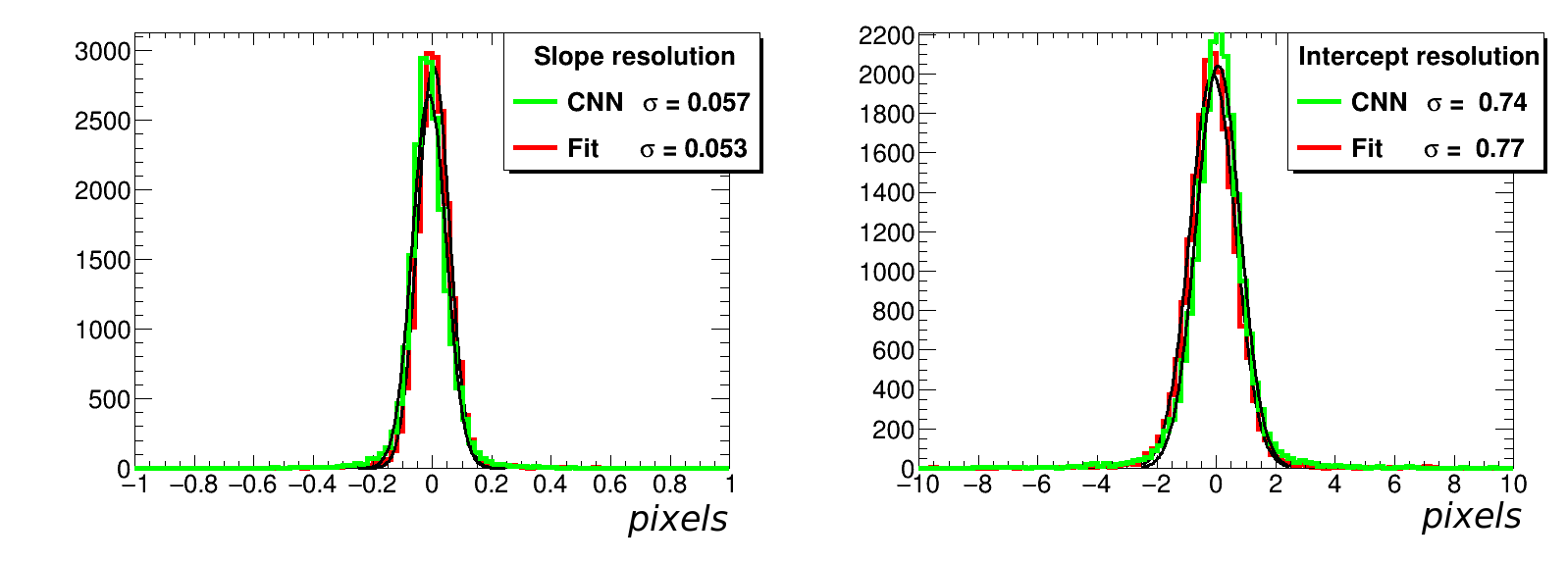}	
	\includegraphics[width=1.0\textwidth]{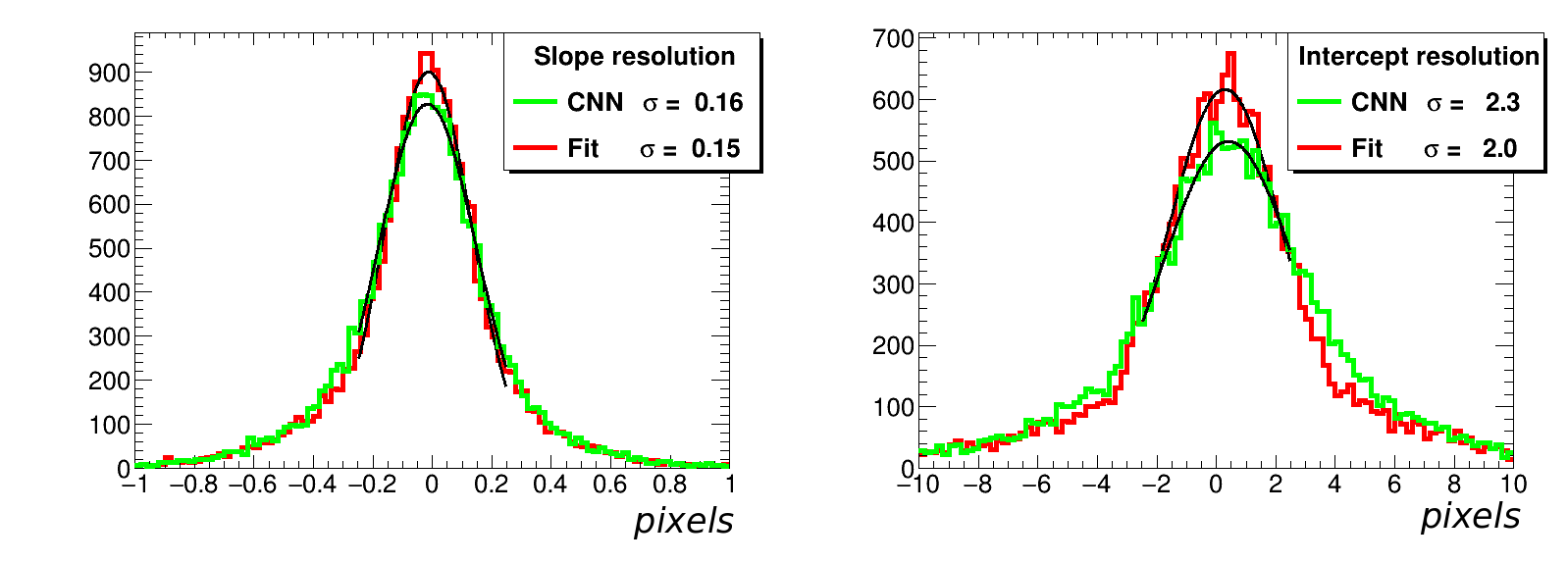}	
	\caption{Resolution of slope (left column) and intercept (right column) for single-track events at pixel efficiency 70\% and noise levels of 10\% (upper row) and 30\% (bottom row).}
	\label{fig_res_single}
\end{figure}

\begin{figure}[!htbp]
	\centering
	\includegraphics[width=1.0\textwidth]{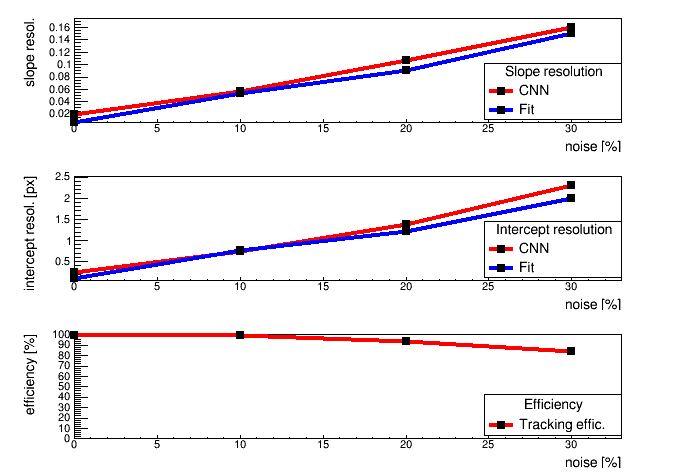}	
	\caption{Track slope resolution obtained by neural network and robust fit (upper plot), track intercept resolution (middle plot), and reconstruction efficiency for single-track events (bottom plot) for noise levels of 0, 10, 20, and 30\%.}
	\label{fig_eff_single}
\end{figure}

Figure~\ref{fig_res_single} shows the distribution of the difference between the reconstructed and true track parameters (slope and intercept) for events with no noise and with a high 30\% noise. The Gaussian is fitted to these distributions; the width of the Gaussian is taken as a measure of the resolution. 
The fit range is chosen to get the width of the central part of the distribution without including the non-Gaussian tails. This is arbitrary and can influence the resolution measurement; however, it is enough to show the trend of how the resolution changes with increasing noise for the toy model described here.
The results for various noise levels are shown in Figures~\ref{fig_res_single} and \ref{fig_eff_single} and summarized in Table~\ref{tab_res_1tr}.

\begin{table}[htb]
	\label{tab_res_1tr}
	\begin{tabular}{|l|c|c|c|c|c|}
		\hline 
		& \multicolumn{2}{c|}{\textbf{ Slope}} & \multicolumn{2}{c|}{\textbf{ Intercept} [pixels]} &
		 \textbf{Efficiency} [\%] \\
		\cline{2-5} 
		& \textbf{CNN} & \textbf{Fit} & \textbf{CNN} & \textbf{Fit} & \\
		\hline 
		\textbf{0\% noise}	& 0.020  & 0.007 & 0.25 & 0.11 & 99\\ 
		\hline 
        \textbf{10\% noise}	& 0.057  & 0.053 & 0.74 & 0.77 & 99\\ 
		\hline 
		\textbf{20\% noise}	& 0.106  & 0.091 & 1.37 & 1.22 & 94 \\ 
		\hline 
        \textbf{30\% noise}	& 0.161  & 0.152 & 2.32 & 2.03 & 84 \\ 		 
		\hline 
	\end{tabular} 
	\caption{Slope and intercept resolution obtained by neural network (CNN) and fit (Fit) together with reconstruction efficiency for single-track events at pixel efficiency of 70\% and different noise levels.}
\end{table} 

The track-reconstruction accuracy is very good for events with no noise; it degrades at higher noise levels. The robust fit applied after the neural network pattern recognition does not significantly improve the resolution. 

The track is assumed to be properly reconstructed when the deep neural network finds a track, which is not further than five pixels from the true track at any point. This cut is arbitrary, but this approach enables us to show how the resolution changes with the increasing noise (which is more important than the exact measurement of the reconstruction efficiency in the toy model). The reconstruction efficiency is defined as the ratio of the number of reconstructed tracks (within the five pixel cut-off) to the total number of generated tracks.
The reconstruction efficiency is measured for the fitted tracks.
For the events without noise, the reconstruction efficiency (shown in Figure~\ref{fig_eff_single}) reaches to above 99\% and drops down to 84\% for 30\% noise.

\subsection{Events with multiple tracks}  

\begin{figure}[!htbp]
	\centering
	\includegraphics[clip,trim={80px 0 100px 0},width=0.40\textwidth]{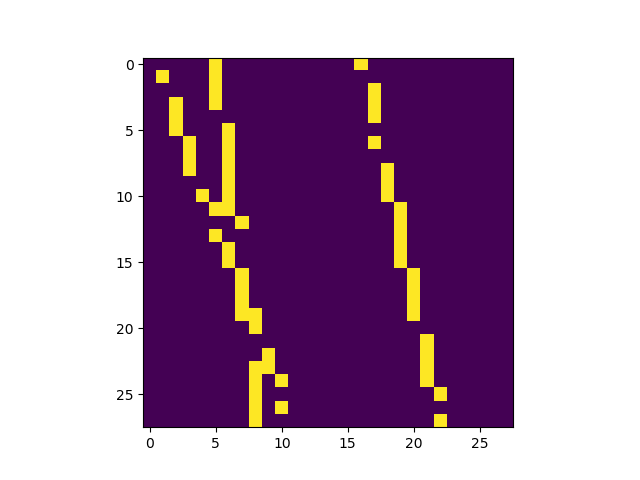}	
    \includegraphics[clip,trim={80px 0 100px 0},width=0.40\textwidth]{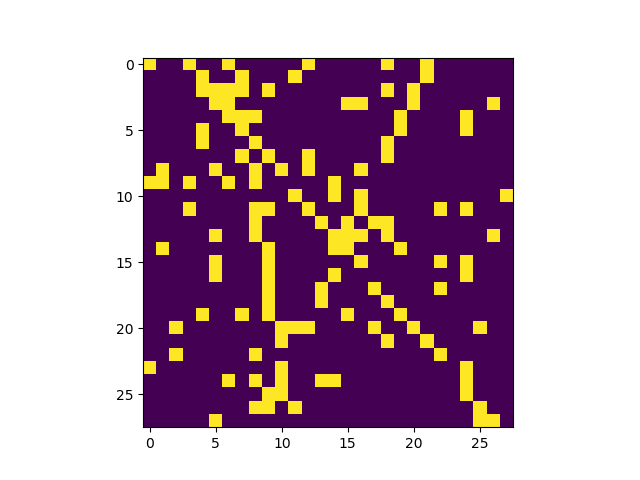}	
    \includegraphics[clip,trim={80px 0 100px 0},width=0.40\textwidth]{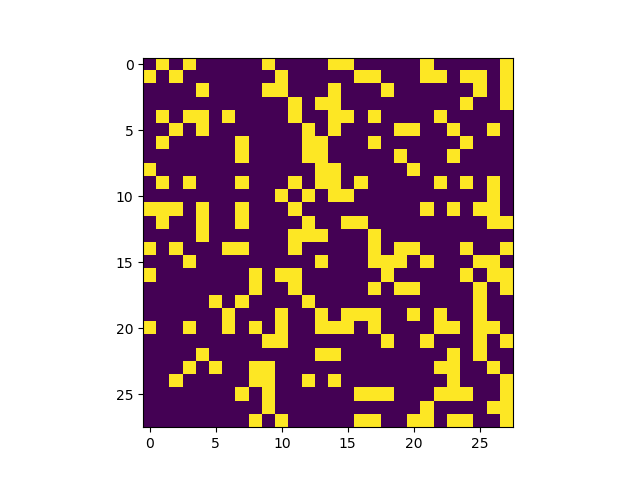}	
    \includegraphics[clip,trim={80px 0 100px 0},width=0.40\textwidth]{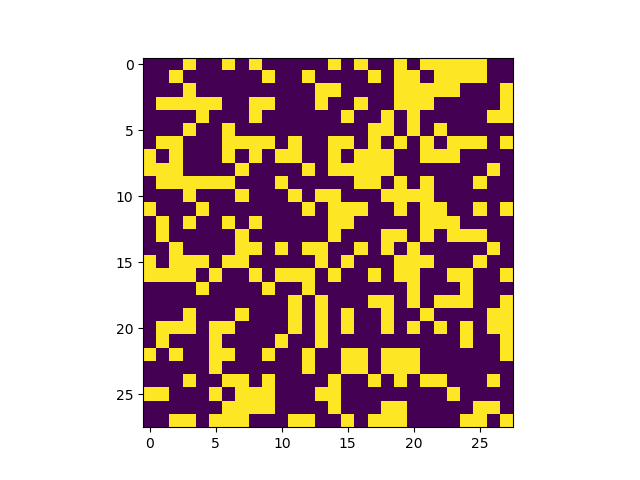}	
	\caption{Three-track events with pixel efficiency of 70\% and noise levels of 0\% (upper left), 10\% (upper right), 20\% (bottom left), and 30\% (bottom right). Units are pixel numbers.}
	\label{fig_multi_event}
\end{figure}

In the next step, the LSTM layer was added to the network as described in Section~\ref{reco}. This layer allows us to reconstruct events with multiple tracks. Example events with three tracks are shown in Figure~\ref{fig_multi_event}.

\begin{figure}[!htbp]
	\centering
	\includegraphics[width=1.0\textwidth]{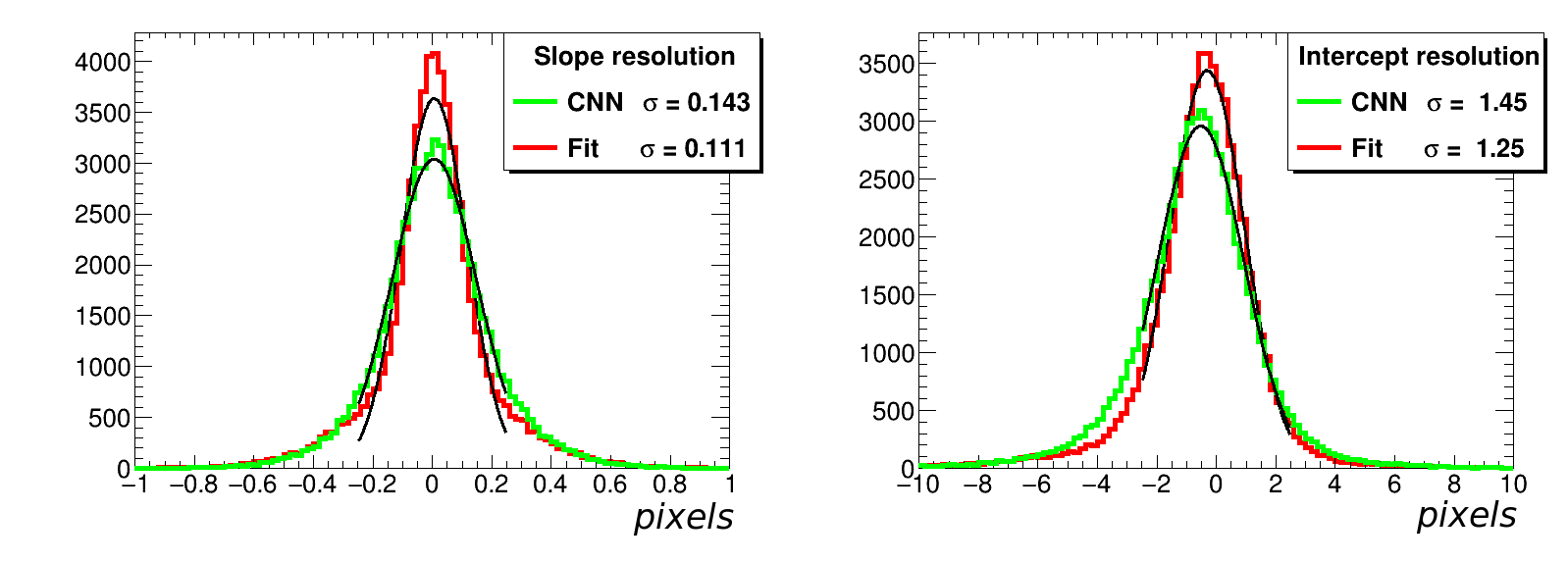}	
    \includegraphics[width=1.0\textwidth]{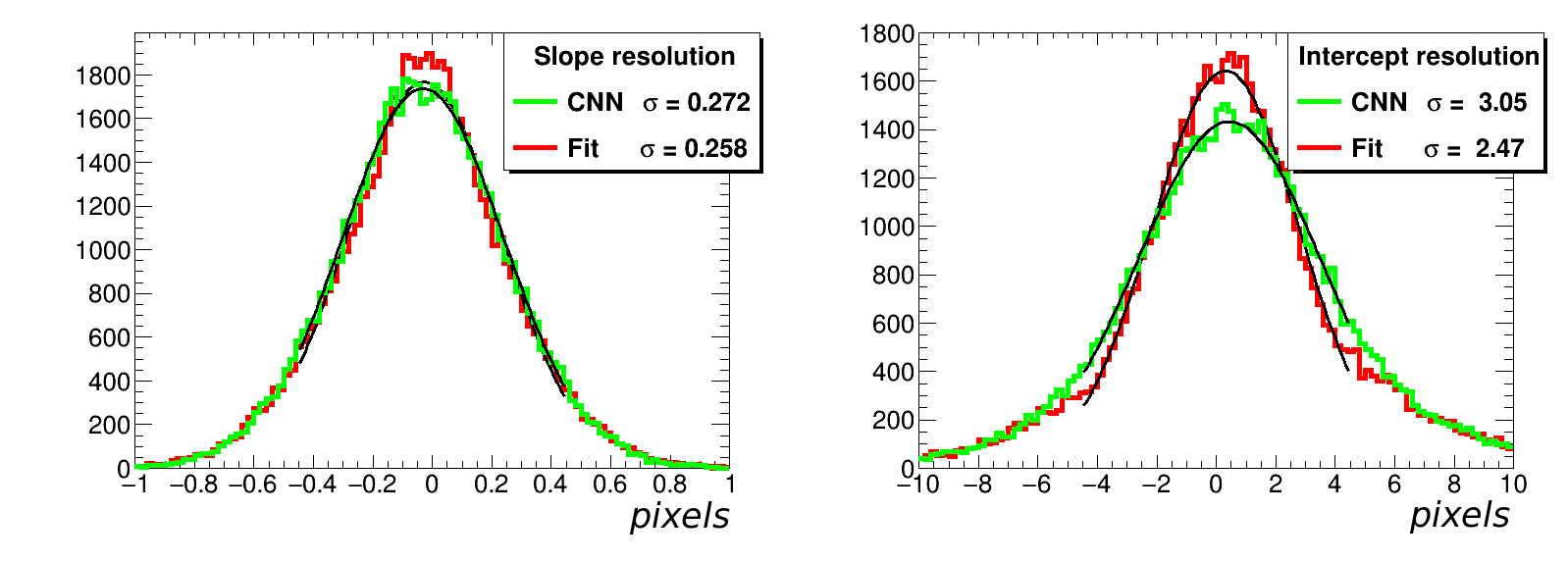}	
	\caption{Resolution of slope (left column) and intercept (right column) for three-track events with pixel efficiency of 70\% and noise levels of 10\% (upper row) and 30\% (bottom row).}
	\label{fig_res_multi}
\end{figure}

\begin{figure}[!htbp]
	\centering
	\includegraphics[width=1.0\textwidth]{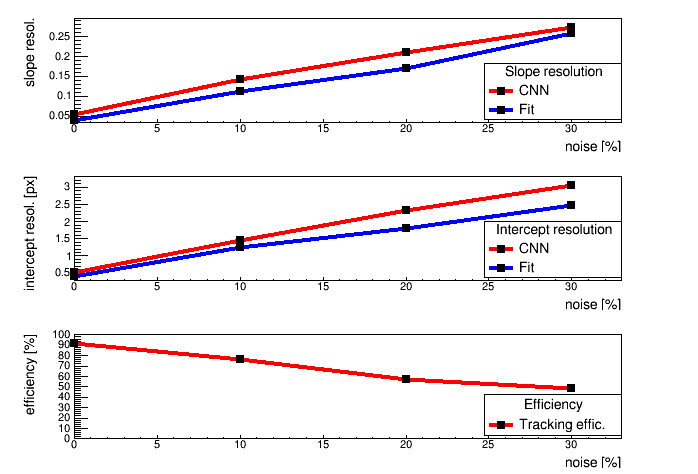}	
	\caption{Track slope resolution obtained by neural network and robust fit (upper plot), track intercept resolution (middle plot), and reconstruction efficiency for events with three tracks (bottom plot) for noise levels of 0, 10, 20, and 30\%.}
	\label{fig_eff_multi}
\end{figure}

Pattern recognition with three tracks at high noise levels is not an easy task, as the tracks are barely visible by the eye. So, it is not surprising that the reconstruction efficiency goes down to about 50\% for 30\% noise. At a 10\% noise level, the efficiency is still reasonable (76\%); for events with no noise, this reaches 92\%. The resolution is significantly worse than for single-track events (see Figures~\ref{fig_res_multi} and~\ref{fig_eff_multi} and Table~\ref{tab_res_3tr}).

\begin{table}[htb]
	\label{tab_res_3tr}
	\begin{tabular}{|l|c|c|c|c|c|}
		\hline 
		& \multicolumn{2}{c|}{\textbf{ Slope}} & \multicolumn{2}{c|}{\textbf{ Intercept} [pixels]} &
		\textbf{Efficiency} [\%] \\
		\cline{2-5} 
		& \textbf{CNN} & \textbf{Fit} & \textbf{CNN} & \textbf{Fit} & \\
		\hline 
		\textbf{0\% noise}	& 0.055  & 0.039 & 0.53 & 0.41 & 92\\ 
		\hline 
		\textbf{10\% noise}	& 0.143  & 0.111 & 1.45 & 1.25 & 76\\ 
		\hline 
		\textbf{20\% noise}	& 0.209  & 0.170 & 2.32 & 1.81 & 57 \\ 
		\hline 
		\textbf{30\% noise}	& 0.272  & 0.258 & 3.05 & 2.47 & 48 \\ 		 
		\hline 
	\end{tabular} 
	\caption{Slope and intercept resolution obtained by neural network (CNN) and fit (Fit) together with reconstruction efficiency for multiple-track events at pixel efficiency of 70\% and different noise levels.}
\end{table} 
%

\begin{figure}[!htbp]
	\centering
	\includegraphics[width=0.48\textwidth]{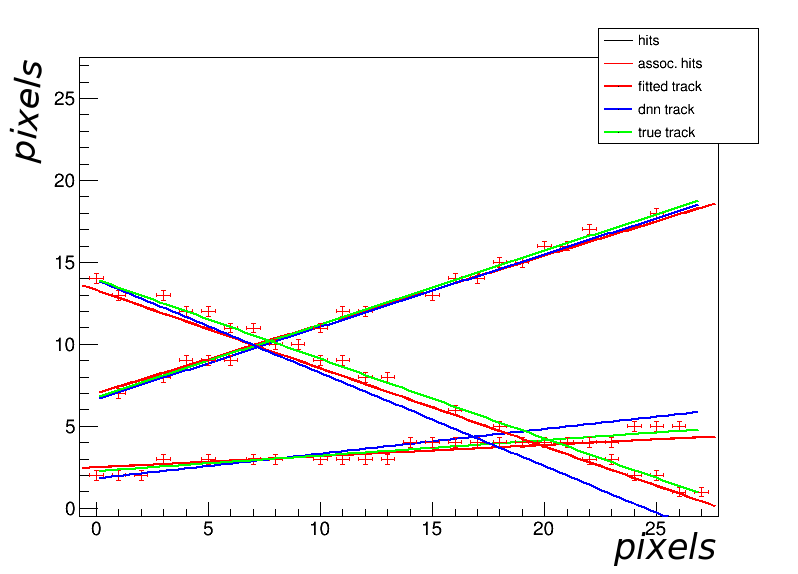}	
	\includegraphics[width=0.48\textwidth]{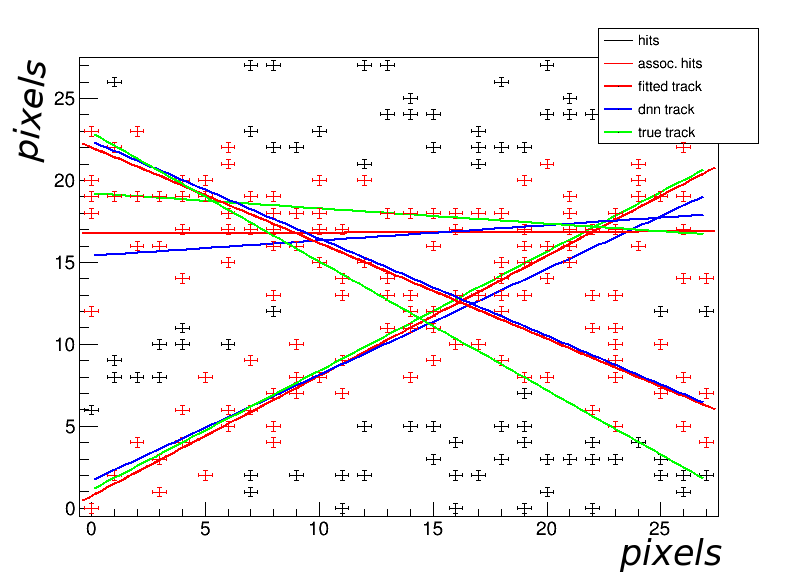}	
	\includegraphics[width=0.48\textwidth]{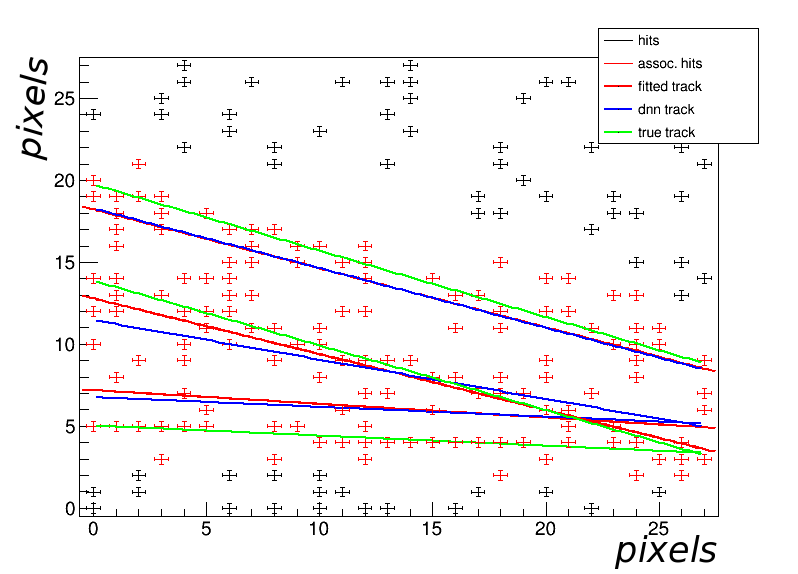}	
	\includegraphics[width=0.48\textwidth]{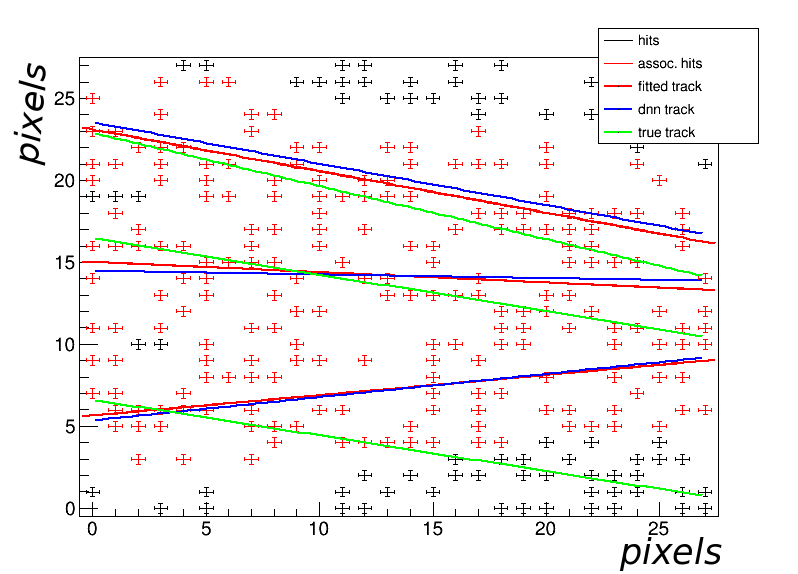}	
	\caption{Example three-track events with tracks found by neural network (blue line) and robust fit (red line) together with true tracks (green line). Hits are denoted as crosses; those marked in red were used for fitting. Events were generated with no noise (upper left plot), 10\% noise (upper right plot), 20\% noise (bottom left plot), and 30\% noise (bottom right plot). Units are pixel numbers.}
	\label{fig_example3t}
\end{figure}

The example events with three tracks and various noise levels are shown in Figure~\ref{fig_example3t}.
The tracks found by the neural network (and afterward by a fit) can be further than five~pixels away from the true track. In this case, they are not counted as ``reconstructed'' for the efficiency measurement. 

For events with higher noise levels, some tracks shown in Figure~\ref{fig_example3t} remain unreconstructed or are reconstructed poorly. However, the high 30\% noise level and relatively low efficiency of 70\% are extreme conditions for the pattern-recognition algorithm.

\section{CPU consumption}

In the toy model presented here, the standard pattern-recognition methods (like the Kalman filter) are not applied; therefore, they cannot be compared to the track finding provided by the neural network. The comparison can be done only between the CPU's time needed to perform the DNN pattern recognition and the CPU used to perform the fit.  
Table~\ref{tab_cpu} shows the wall-clock time on our PC needed for the network training, the actual pattern recognition, and the following fit for the three-track events and 10\% noise. 
\begin{table}[H]
\label{tab_cpu}
 \begin{tabular}{|l|r|}
	\hline 
	& \textbf{Wall-time [sec]}  \\ 
	\hline 
\textbf{Neural network training}	& 417 \\ 
	\hline 
\textbf{Neural network pattern recognition}	&  10 \\ 
	\hline 
\textbf{ROOT robust fit}	& 467 \\ 
	\hline 
 \end{tabular} 
\caption{CPU consumption (wall-clock time on 4-core PC) by neural network pattern recognition and robust fit for three-track events with 10\% noise (data samples of 20,000 events used for training and analysis)}
\end{table} 
%
%
%

The Tensorflow package in which the DNN is implemented is a multi-threaded application and can use all four cores of our PC. The fit implemented in ROOT can use only a single core.

The neural network requires a lot of CPU time for training; however, this time is needed only once (before starting the actual analysis). This could be further shortened if graphics processor units (GPUs) were used. Once the network is trained, the pattern recognition needs about a 50-times-lower wall-clock time than the robust fit. This corresponds to about a 12-times-lower CPU time, since the fit uses a single core only.

\section{Future improvements of tracking algorithm}

The most obvious extension of the presented track-finding algorithm is the reconstruction of tracks in three dimensions. This could be done by replacing the 2D convolutional layers in a neural network with 3D  layers.
The size of the network should be also increased to accomodate more complex 3D input events.

The more important improvement would be the use of the mixture density network (MDN) already proposed in 1994~\cite{bishop1994mixture}. In this approach, the output of the network in the case of regression is not a vector of the most probable outputs for a given input vector but a probability density of the possible outputs.

This approach makes MDNs similar to Bayesian Neural Networks (BNN)~\cite{pearl1988markov,neal2012bayesian}, which also return the probability density distribution rather than the actual most probable output vectors. The advantage of MDNs is that, in contrast to BNNs, they do not require the training of multiple neural networks; this makes them much faster and easier to train and apply. 

In the case of track reconstruction, 
the obvious advantage of MDN is that it learns the probability distribution, therefore learning both the track position and its uncertainty at the same time in a single network. The network then returns the uncertainties of the track parameters, allowing only for the selection of well-reconstructed tracks and improving the performance of the track reconstruction. 

The less obvious advantage for track finding is that the usage of the mixture density network (MDN) makes the long/short-term memory (LSTM) layer no longer necessary, therefore simplifying the network. The tracks do not need to be ordered like in the case of LSTM (which was always slightly artificial). In the MDN approach, a single Gaussian might be used to describe the probability density of each parameter of a track, and the number of Gaussians sets the limit on the number of tracks. Therefore, the network still stays flexible, being able to reconstruct any number of tracks below this limit.

\section{Application in future experiments}

As the applied algorithms have been proven to be successful on a toy Monte Carlo model with three tracks, they are planned to be used in more-realistic conditions using testbeam data for the MUonE experiment planned to be operating at the SPS accelerator at CERN~\cite{MUonE}. This is characterized by a relatively low occupancy; however, due to the limited acceptance and very low angles of the outgoing particles, a separation between an electron and a muon after elastic scattering is not straightforward and needs efficient and very precise track-reconstruction algorithms. Therefore, in order to achieve the physics goal (which is a measurement of the hadronic contribution to the anomalous muon magnetic moment with an uncertainty at a level of 0.05\%), the deep neural network methods are going to be applied and tested with respect to the commonly used pattern-recognition and tracking procedures based on histogramming or simple line fitting. 

\begin{figure}[hbt]
	\centerline{
		\includegraphics[scale=0.60]{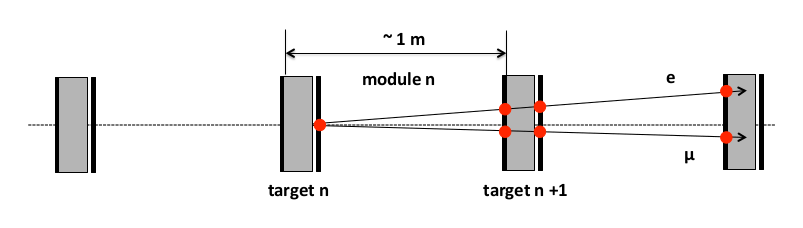}}
	\caption{Concept of MUonE detector's final layout, where input beam enters detector from left. Thick black lines denote tracking modules. Figure adopted from~\cite{MUonE}.}
	\label{fig:MUonEDet}
\end{figure}

During the CERN 2018 run, the MUonE experiment performed a feasibility test at the COMPASS experiment site, collecting a data sample size of about 1.4 billion events. Figure~\ref{fig:MUonEDet} shows the concept of the MUonE setup; it is a modular system consisting of 16 silicon micro-strip layers, 2~carbon targets, and an electromagnetic calorimeter. The modules are separated by a relative distance of 1~m from each other and spaced by air.

There are three tracks to be reconstructed in the detector; i.e., the direction and momentum of the incident muon as well as the directions of the outgoing electron and muon. Therefore, according to no CPU time limit and the expected relatively low detector occupancy, the goal of the pattern recognition is to maximize the possible track reconstruction efficiency. Standard pattern-recognition and track-finding procedures were applied, and the obtained results are going to be compared with the outcome of the parallel deep neural network method (which is a natural candidate for further improvements in the reconstruction performance). In principle, this allows for performing the pattern recognition ``at once'' while at the same time having great potential for increasing the precision and efficiency.

\section{Conclusions}

The neural network algorithm was proven to be successful when applied to the toy model. 
However, it is difficult to make any judgment on whether the network described in this article can be successfully used in a real experiment for track reconstruction. This requires more studies and further development of the tracking system. 

In the next step, we plan to test the track reconstruction in more-realistic conditions using testbeam data taken by the MUonE experiment. Therefore, in order to reduce the CPU timing, increasing the efficiency of the overall tracking reconstruction procedure and improving the precision the deep neural network methods are going to be applied and tested with respect to the commonly used procedures. Such tests could give an answer for whether pattern recognition based on deep neural networks can compete with standard track-finding procedures.

The toy model described in this article does not require much computing power; however, if extended to real-size events and to three dimensions, more-powerful machines or a GPU cluster would be useful for speeding up the network training.

\begin{acknowledgements}
The work was supported in part by 
	Polish Government Grant No. 2017/25/B/ST2/01234
	and ACK CYFRONET Computing Grant No. lhcbflav08.

\end{acknowledgements}

\nocite{*}

\bibliographystyle{cs-agh}
\bibliography{bibliography}

\begin{thebibliography}{10}
\providecommand{\url}[1]{\texttt{#1}}
\providecommand{\urlprefix}{URL }

\bibitem{LSTMtutorial}
Keras LSTM tutorial.
\newblock \urlprefix\url{https://adventuresinmachinelearning.com/keras-lstm-tutorial/}.

\bibitem{abadi2016tensorflow}
Abadi M., Barham P., Chen J., Chen Z., Davis A., Dean J., Devin M., Ghemawat S., Irving G., Isard M., et~al.: Tensorflow: A system for large-scale machine learning.
\newblock In: \emph{12th $\{$USENIX$\}$ Symposium on Operating Systems Design and Implementation ($\{$OSDI$\}$ 16)}, pp. 265--283. 2016.

\bibitem{MUonE}
Abbiendi G., al.: Measuring the leading hadronic contribution to the muon $g-2$ via $\mu e$ scattering.
\newblock In: \emph{European Physical Journal C}, vol.~77, p. 139, 2017.

\bibitem{bishop1994mixture}
Bishop C.M.: Mixture density networks.
\newblock In: , 1994.

\bibitem{Brun:1997pa}
Brun R., Rademakers F.: {ROOT: An object oriented data analysis framework}.
\newblock In: \emph{Nucl. Instrum. Meth.}, vol. A389, pp. 81--86, 1997.
\newblock \urlprefix\url{http://dx.doi.org/10.1016/S0168-9002(97)00048-X}.

\bibitem{heptrkx}
Calafiura P.: HEP advanced tracking algorithms with cross-cutting applications (Project HEP.TrkX).
\newblock \urlprefix\url{https://heptrkx.github.io/}.

\bibitem{chollet2015keras}
Chollet F., et~al.: Keras: Deep learning library for theano and tensorflow, 2015.
\newblock \urlprefix\url{https://keras.io/}.

\bibitem{farrell2017hep}
Farrell S., Anderson D., Calafiura P., Cerati G., Gray L., Kowalkowski J., Mudigonda M., Spentzouris P., Spiropoulou M., Tsaris A., et~al.: The HEP. TrkX Project: deep neural networks for HL-LHC online and offline tracking.
\newblock In: \emph{EPJ Web of Conferences}, vol. 150, p. 00003. EDP Sciences, 2017.

\bibitem{farrell2018novel}
Farrell S., Calafiura P., Mudigonda M., Anderson D., Vlimant J.R., Zheng S., Bendavid J., Spiropulu M., Cerati G., Gray L., et~al.: Novel deep learning methods for track reconstruction.
\newblock In: \emph{arXiv preprint arXiv:1810.06111}, 2018.

\bibitem{fruhwirth1987application}
Fr{\"u}hwirth R.: Application of Kalman filtering to track and vertex fitting.
\newblock In: \emph{Nuclear Instruments and Methods in Physics Research Section A: Accelerators, Spectrometers, Detectors and Associated Equipment}, vol. 262(2-3), pp. 444--450, 1987.

\bibitem{graves2014neural}
Graves A., Wayne G., Danihelka I.: Neural Turing Machines.
\newblock In: \emph{arXiv preprint arXiv:1410.5401}, 2014.

\bibitem{hochreiter1997long}
Hochreiter S., Schmidhuber J.: Long short-term memory.
\newblock In: \emph{Neural computation}, vol.~9(8), pp. 1735--1780, 1997.

\bibitem{james1975minuit}
James F., Roos M.: MINUIT: a system for function minimization and analysis of the parameter errors and corrections.
\newblock In: \emph{Comput. Phys. Commun.}, vol.~10(CERN-DD-75-20), pp. 343--367, 1975.

\bibitem{kingma2014adam}
Kingma D.P., Ba J.: Adam: A method for stochastic optimization.
\newblock In: \emph{arXiv preprint arXiv:1412.6980}, 2014.

\bibitem{lecun2015deep}
LeCun Y., Bengio Y., Hinton G.: Deep learning.
\newblock In: \emph{nature}, vol. 521(7553), p. 436, 2015.

\bibitem{motulsky2006detecting}
Motulsky H.J., Brown R.E.: Detecting outliers when fitting data with nonlinear regression--a new method based on robust nonlinear regression and the false discovery rate.
\newblock In: \emph{BMC bioinformatics}, vol.~7(1), p. 123, 2006.

\bibitem{neal2012bayesian}
Neal R.M.: \emph{Bayesian learning for neural networks}, vol. 118.
\newblock Springer Science \& Business Media, 2012.

\bibitem{pearl1988markov}
Pearl J.: Markov and Bayesian Networks, chap. 3 \emph{Probabilistic Reasoning in Intelligent Systems}, 1988.

\bibitem{srivastava2014dropout}
Srivastava N., Hinton G., Krizhevsky A., Sutskever I., Salakhutdinov R.: Dropout: a simple way to prevent neural networks from overfitting.
\newblock In: \emph{The journal of machine learning research}, vol.~15(1), pp. 1929--1958, 2014.

\bibitem{vinyals2015show}
Vinyals O., Toshev A., Bengio S., Erhan D.: Show and tell: A neural image caption generator.
\newblock In: \emph{Proceedings of the IEEE conference on computer vision and pattern recognition}, pp. 3156--3164. 2015.

\end{thebibliography}

\end{document}